\documentclass[conference, twocolumn]{IEEEtran}
\usepackage{mdwlist}
\usepackage{cite}

\usepackage{amsmath}
\usepackage{amssymb}
\usepackage{amsthm}

\usepackage[final]{graphicx}
\usepackage{color}
\usepackage{subfigure}
\usepackage{epstopdf}

\usepackage{wrapfig}
\usepackage{hyperref}
\hypersetup{
	colorlinks=true
}

\newtheorem{theorem}{Theorem}[section]

\newtheorem{lemma}[theorem]{Lemma}

\theoremstyle{definition}

\theoremstyle{remark}
\newtheorem{remark}[theorem]{Remark}

\title{Game Theoretic Formation of a Centrality Based Network}
\author{
\IEEEauthorblockN{Ryan Tatko}
\IEEEauthorblockA{Applied Research Laboratory \&\\
Department of Economics\\
Penn State University\\
University Park, PA 16802\\
E-mail: \texttt{rwt5069@psu.edu}}
\and
\IEEEauthorblockN{Christopher Griffin}
\IEEEauthorblockA{Applied Research Laboratory \&\\
Department of Mathematics\\
Penn State University\\
University Park, PA 16802\\
E-mail: \texttt{griffinch@ieee.org}}
}

\begin{document}
\maketitle

\begin{abstract} 
We model the formation of networks as a game where players aspire to maximize their own centrality by increasing the number of other players to which they are path-wise connected, while simultaneously incurring a cost for each added adjacent edge.  We simulate the interactions between players using an algorithm that factors in rational strategic behavior based on a common objective function.  The resulting networks exhibit pairwise stability, from which we derive necessary stable conditions for specific graph topologies.  We then expand the model to simulate non-trivial games with large numbers of players.  We show that using conditions necessary for the stability of star topologies we can induce the formation of hub players that positively impact the total welfare of the network.
\end{abstract}

\section{Introduction}

In the last fifteen years, the emerging study of network science has produced results impacting a broad variety of dynamic systems, from biological growth to human interaction \cite{BB09,JW02}.  The recent surge in society's dependence on social networks places extra importance in the study of how these networks form, grow, and eventually stagnate.  Although commonly accepted models produce some accurate network characteristics, they rely almost purely on probabilistic methods of network growth \cite{Alderson2006}.  Even the most popular stochastic generative models like Watts and Strogatz's small-world \cite{WS99} and Barab\'{a}si and Albert's preferential attachment models \cite{BA99} fall short of describing real network behavior.  In contrast, we will explain the formation of networks using game theoretic principles, where individual ``players" make rational decisions based on maximizing an assigned payoff function under a given information set.  Applying game theory allows us to model social behavior and interpret the resulting network structure using the underlying economic factors that a player might take into consideration.  

In this paper, we study the dynamic formation of complex networks where the principle source of utility for each player comes from how central he is in the network.  The concept of \textit{centrality}, as a metric, is commonly used in graph theory and social network analysis to measure the relative importance of a single node within a graph \cite{CSW05,KY08}.  Consequently, centrality can be applied to networks to gauge how influential an individual is, determine the chain of command within an organization, or in theory, study how social constructions such as cities form and develop.  Individuals often place importance on being in the ``middle"; those with high centrality in a social network often learn about new information before most of the rest of the group.

There are numerous measures of centrality commonly used in network analysis; betweenness, closeness, degree centrality \cite{New10}, and even Google's PageRank \cite{BP98,H03} are each used to explain centrality in different contexts.  In the case of degree centrality, nodes with a higher number of direct connections relative to other nodes in a graph are more central to the ``flow" of information, as the (unweighted) shortest paths through the graph tend to traverse through highly central nodes rather than less central ones.  While the number of direct links is a large factor in an node's centrality, the number of \textit{indirect} links to a node is also important in determining the influence of a specific node within a network.  When we consider indirect connectivity, degree centrality is not sufficient.  

	Introduced by Leo Katz in the 1950's \cite{KZ53}, Katz centrality is a generalization of degree centrality that measures the relative influence of a node within a network. The Katz centrality of a node primarily depends on that node's immediate neighbors, but also the nodes connected to these immediate neighbors and so on.  Formally, the Katz centrality of node $i$ is defined as
\begin{equation}
	C_{K_{i}} = \sum_{k=1}^\infty \sum_{j=1}^n \alpha^k (\mathbf{A}^k)_{ij} \qquad \alpha\in(0,1)
\end{equation}

Where $\mathbf{A}$ is the adjacency matrix of the network.  Here, the powers of $k$ measure the presence of links through intermediate nodes.  For example, in the matrix $\mathbf{A}^2$, if $a_{13}$ = 1, node $1$ and node $3$ are connected through one immediate neighbor.  Distant links are penalized by a factor $\alpha \in (0,1)$ that assigns weights to each link based on the distance between the nodes.  For this definition of centrality to be meaningful, $\alpha$ must be smaller than $1/\lambda_{0}$ where $\lambda_{0}$ is the largest eigenvalue of the adjacency matrix $\mathbf{A}$.  Assuming this, Katz centrality can be calculated as
\begin{equation}
	\mathbf{C}_{K} = ((I - \alpha \mathbf{A}^T)^{-1} - I)\mathbf{1}_{n} \label{vecC}
\end{equation}

Here $\mathbf{A}^T$ is the transposed adjacency matrix of the network, $\mathbf{I}_{n}$ is the $n \times n$ identity matrix, and $\mathbf{1}_{n}$ is a ones vector of size $n$. We use this form of centrality in order to apply our results to directed graphs as well as undirected graphs.

\section{Literature review}

The study of networks is a relatively new direction in the game theory and network science literature \cite{GS12,SSG11,SG12a}.  Jackson and Wolinski \cite{JW96} analyzed the relationship between stability and efficiency of simple economic networks composed of individuals equipped with a utility function.  From this, they were able to deduce network topologies most likely to form given a specific set of conditions related to redistributive structures of the network.  More recent work by Goyal and Joshi \cite{GJ03} looked at networks resulting from the formation of oligopolies between collaborating firms.  Their research focused on firms forming $pairwise$ $stable$ links, a key characteristic of networks that depend on a mutual benefit to maintain a connection. Additional work on this is extensive. See \cite{JW96,dutta1997,bala2000,furusawa2007,belleflamme2004,calvo2007,jackson2003} and their references. In this paper we analyze different topologies of pairwise stable networks and the dynamics of heterogeneous link costs between players.
     
\section{Notational Preliminaries}

	In this section we introduce conventional network science and game theory notation.  Assume that there is a set of players, $N=\{1,2,3,\dots,n\}$ where $n \in \mathbb{N}$.  A graph $G$ is a defined by a set of vertices (players) and edges connecting them.  If we define $G^N$ to be the complete set of all possible adjacent edges, the set of all graphs over $N$ is defined as $G'=\{G:G \subset G^N\}$.  The set of $N$ players within a network have relationships characterized by binary variables, where $a_{ij}\in\{0,1\}$ represents this relationship between any two players $i,j$.  Let $a_{ij}=1$ if $i$ is directly connected to $j$ and 0 otherwise.  The \textit{degree} of a player $\eta_{i}$ is the number of direct connections $i$ has in the graph, defined as $\sum_{j=1}^na_{ij}$.  As is common in network science literature, we ignore the nonsensical possibility of self loops so that $a_{ii}$=0.  The complete graph $K_{n}$ is a graph which is defined by a specific degree sequence $\eta_{i}=n-1$, $\forall\, i \in N$.
	
In our network formation game, we define a simple strategy set $S_{i}=\{s_{i}^1,s_{i}^2\}$ that gives each player the ability to form ($s_{i}^1$) or veto ($s_{i}^2$) a link with another player.  We equate the act of vetoing a link with deleting a pre-existing connection between two players, if it exists.  Let $S=\Pi_{i=1}^n S_{i}$ be the set of all possible strategy profiles.  We define a network game $\mathcal{G}(N, S, \pi)$, where $\pi:S\to\mathbb{R}$ is a payoff function assigned to the set of players $N$.

\section{Theoretical Results}

\subsection{Objective Model}

In our work, we consider a non-cooperative game structure where a player's objective relies on maximizing his relative centrality while simultaneously increasing the size of the network.  In the context of real world networks, highly centralized individuals often possess the largest ``sphere of influence," which becomes most powerful when in close proximity to a vast number of people.  Specifically, we will study the case where players attempt to maximize the probability they are landed on by a random walk within their connected component.  We can capture this objective as:

\begin{equation}
	\lambda_{i} = R_{i}p_{i}K_{i}
\end{equation}

Here $\lambda_{i}$ is the benefit function equal to the number of nodes in $i$'s component (minus itself) $p_{i}$, multiplied by the scaled component-wise Katz centrality of the node, $K_{i}$.  This benefit function captures an interesting trade off the player faces between maintaining a high centrality while increasing the size of his component.  We then multiply this value by some arbitrary award $R_i$ a player (individuals, store, websites, etc.) receives when another player (potentially not directly connected) engages in an interaction with this vertex. The centrality in the payoff function is defined as:

\begin{equation}  
	K_{i} = \frac{C_{K_{i}}}{\sum_{j \in H(j)}C_{K_{j}}}
\end{equation}

Where $C_{K_{i}}$ is equal to the Katz centrality of $i$ in the player's connected subgraph $H(i)$.  Our centrality measurement acts as a probability mass that maintains a sum of one no matter the size of the network, and so is consistent with the notion that (all else equal) players lose centrality as the total size of the connected network grows.  To maintain consistency in the model, we define the component centrality of an isolate node to be equal to one.  We may assume a linear cost associated with establishing out-links:

\begin{equation}
	\phi_{i} =\sum_{j\neq i}\gamma_{ij}A_{ij}
\end{equation}

We assume that $\gamma_{ij}=\gamma_{i}$, $\forall j \in N$ and $\gamma_{i}$ is a bilateral fixed cost for establishing individual links for all players.  The payoff function is derived as

\begin{equation} \label{pi}
	\pi_{i} = \lambda_{i}-\phi_{i}=R_{i}p_{i}K_{i}-\sum_{j \neq i}\gamma_{i}A_{ij}
\end{equation}

From this point on, when we denote the link cost as $\gamma$ without subscript we assume that all players share a link cost, so that $\gamma_i = \gamma$ for $i = 1,\dots, N$.  Simple marginal analysis shows that a player $i$ will establish a link with another player $j$ as long as $\Delta\lambda_{i} > \gamma$, or the increase in the benefit for $i$ is greater than the linear cost associated with adding an additional link.  For an undirected graph, a link will be made between $i$ and $j$ if and only if $\Delta \pi_{i},\Delta\pi_{j} > 0$.  In this sense, players attempt to minimize link maintenance cost while simultaneously gaining centrality by establishing (or deleting) links with other players.

\subsection{Stability of Complete Graph Topologies}

Pairwise stability as described by Jackson \cite{JW96} is simply defined as a network where no player benefit from creating a new link, and no two players benefit from severing an existing link. We assume that both players must bilaterally agree to the creation of a link, while any player can sever a link. This is similar to ``friending'' on Facebook.  Under topologically-specific conditions, complete networks and star networks are pairwise stable using our model.  In the following sections, we construct the Katz centrality in manipulatable terms to derive pairwise stable conditions for these network structures.  

By calculating the Katz centrality as defined in \eqref{vecC} explicitly, it is possible to express $\mathbf{K}_{i}$ in terms of the parameters $n$ and $\alpha$.  From this expression, pairwise stability can be defined for $any$ complete graph $K_{n}$ and link cost $\gamma$. Lemma \ref{lem:KnLemma} summarizes this result.

\begin{lemma}
For a complete graph $K_{n}$, the centrality vector $\mathbf{K}$ is given by:
\begin{equation} \label{kn}
		K_{i} = \frac{(n-1)\alpha}{1-(n-1)\alpha} \cdot \frac{1-(n-1)\alpha}{n(n-1)\alpha} = \frac{1}{n}
	\end{equation} 
\label{lem:KnLemma}
	\hfill\qed
\end{lemma}

The expression in Equation $\eqref{kn}$ defines the Katz centrality of a vertex in a complete graph and shows that it is equivalent to the sum of the infinite series $\sum_{k=1}^{\infty}((n-1)\alpha)^k$.  We observe that the each vertex is equally central in the complete graph and the relative centrality of a vertex decreases harmonically as $n$ increases.

To prove stability, we consider the deletion of a link as the one possible complete graph manipulation.  We define a nearly-complete graph as the graph $K_{n}^{(-1)}$ with degree sequence $\{n-1,n-1,\ldots,n-2,n-2\}$.  The centrality of a nearly-complete graph can be derived from $\eqref{vecC}$ by changing the adjacency matrix so that $a_{ij}$ = 0 for exactly one pair of $i,j \in N$.

\begin{lemma}
For a complete graph minus one link between two vertices, the centrality vector $\mathbf{K}_n^{(-1)}$ has two distinct values $K_{b},K_{s}$ given by:
\begin{align} \label{ks}
		&K_{b} = \frac{(2n-2)\alpha+n}{(n-2)(2n\alpha+n+1)}  \nonumber \\
		&K_{s} = \frac{2\alpha+1}{2n\alpha+n+1}
	\end{align}\hfill\qed
\label{lem:KnLemma2}
\end{lemma}

\begin{figure}[htbp]
\centering
\includegraphics[scale = .33]{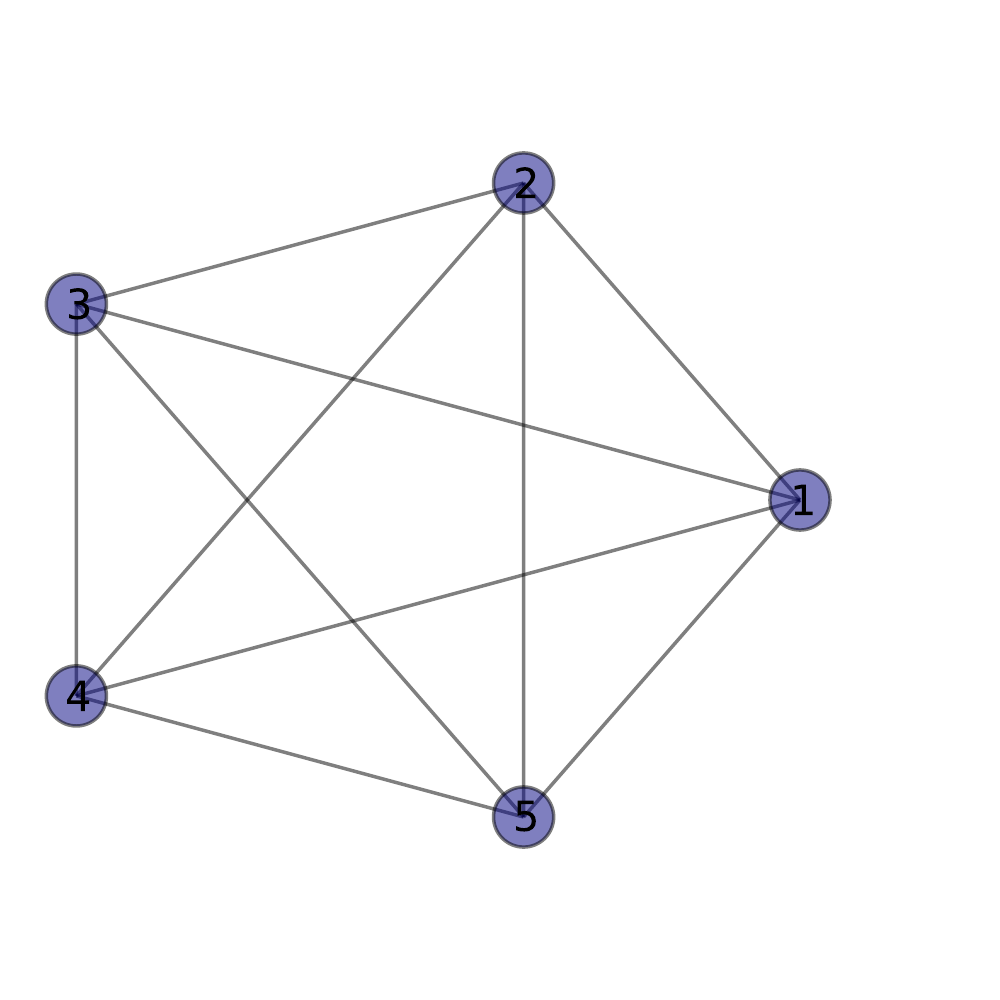}
\includegraphics[scale = .33]{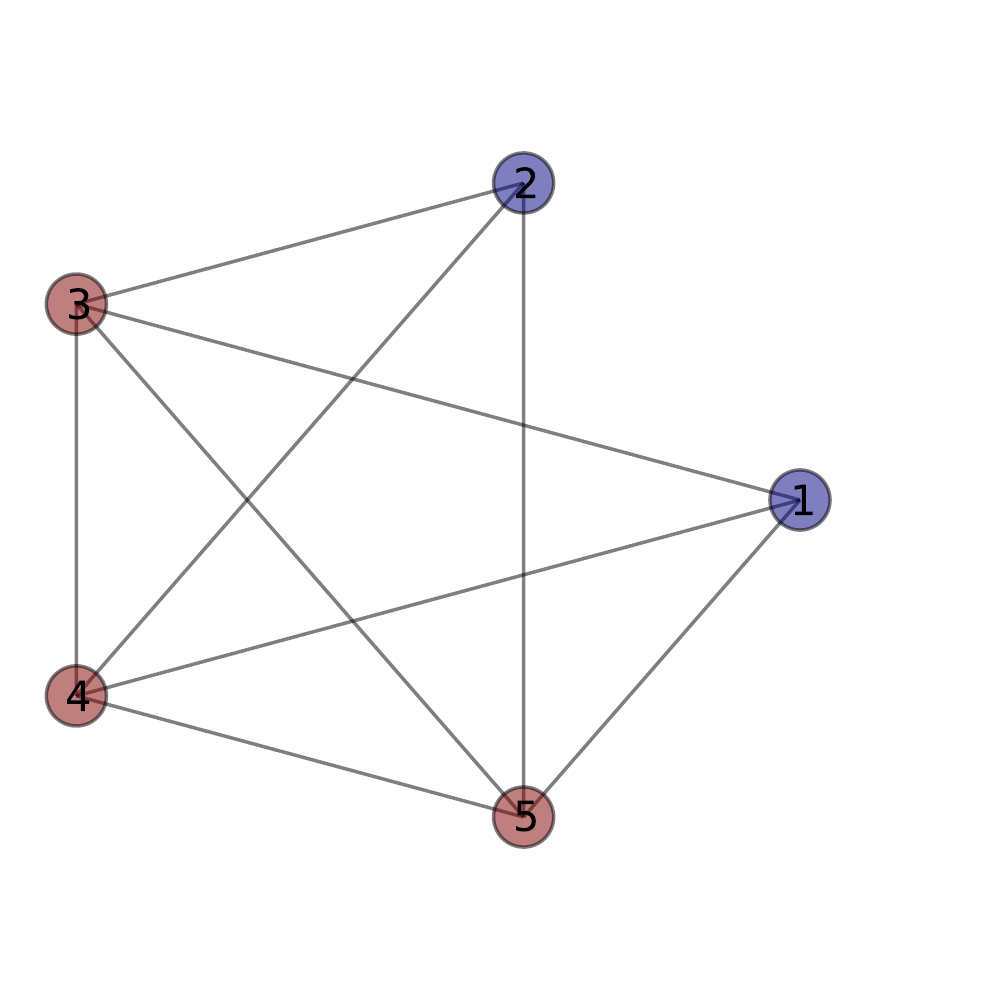}
\caption{The complete graph K$_{5}$ and the nearly complete graph G$_{5,9}$ s.t. $a_{1,2}$=0.  For high values of $\gamma$, one player benefits from a link deletion and the complete graph loses stability.  This causes a cascading effect of s$_{2}$ strategy dominance, resulting in a graph with fewer (and often zero) links.}
\label{fig:KnLemma2}
\end{figure}

Lemma \ref{lem:KnLemma2} is illustrated in Figure \ref{fig:KnLemma2}, where we show the two possible types of vertices in $\mathbf{K}_n^{(-1)}$.

Expression $\eqref{ks}$ shows that there are two possible centrality values for vertices in a nearly-complete graph, where  $K_{s}$ is the centrality of the two vertices with $n-2$ degrees and $K_{b}$ is the centrality for all other vertices.  From the expressions it follows that the deletion of a single link in a complete graph always results in a lower centrality and $K_{s} < K_{b}$.

\begin{theorem}
	A complete network in which $ a_{ij} = 1, \forall i \neq j \in N$ will result in a pairwise stable equilibrium as long as
	$R_{i}(n-1)(K_{i}-K_{i}^{(-1)}) > \gamma$.	
\end{theorem}

\begin{IEEEproof}
	Consider a complete graph $K_{n}$ where $N$ is the set of $n$ players.  By definition, the graph is a single component so $p_{i} = (n-1)$, $\forall i \in N$.  For the complete graph, the payoff function can be simplified as
	\begin{equation}
		\pi_{i}^\star = R_{i}(n-1)K_{i} - (n-1)\gamma \label{pistar}
	\end{equation}
	Where $\pi_{i}^\star$ represents the payoff for each player $i$ in the complete graph $K_{n}$.  If a player deviates away from $\pi_{i}^\star$ by deleting an existing link, that player's payoff is given by
	\begin{equation}
			\pi_{i} = R_{i}(n-1)K_{i}^{(-1)} - (n-2)\gamma
	\end{equation}
	If $\pi_{i} \leq \pi_{i}^\star$, than no player in $K_{n}$ will benefit from deviating from the complete structure.  Thus
	\begin{equation}
		\pi_{i}^\star - \pi_{i} > 0 \implies 
	 	R_{i}(n-1)(K_{i}-K_{i}^{(-1)}) > \gamma \label{gammaineq}
	\end{equation}
From $\eqref{kn}$ and $\eqref{ks}$, we get the explicit necessary condition for stability:
\begin{equation}
\frac{n-1}{n(2n\alpha+n+1)} > \frac{\gamma}{R_{i}}
\end{equation}
As long as this inequality holds, the complete graph is pairwise stable and any player will receive a non-positive marginal payoff from deleting an existing link with another player.
\end{IEEEproof}

Note that if a fixed link cost is chosen and players are subsequently added to the complete network, the necessary stability inequality is violated and the complete network will collapse and reform into smaller components.

\subsection{Stability of Star Graph Topologies}

When we study real world networks, it is difficult to find ``completeness'' on a large scale; rarely does every individual in even the closest communities form bonds with every other individual.  Networks with star topologies are a common phenomena, appearing often in computer and social systems aiming to optimize efficiency.  In its basic form, a star graph $S_{n}$ is defined by $n$ vertices and $n$+1 edges, where each ``leaf" vertex is connected to only a single ``hub" vertex.  We induce the stability in star networks by choosing separate link costs $\delta$ and $\zeta$ for leaf and hub vertices.  These ``tax breaks" simulate a heterogeneous population of players found in nearly every real world system.  These non-trivial networks exhibit interesting properties explained by their stability conditions.

Like our previous section on complete graphs, we derive star graph centrality explicitly in terms of parameters $n$ and $\alpha$.  

\begin{lemma} \label{stthm}
For any star graph $S_{n}$, the centrality vector $\mathbf{K}$ has two possible values: $K_{b}$, the centrality of the central vertex and $K_{s}$, the centrality of the leaves, where:
\begin{align} \label{231}
		K_{b} & = \frac{\alpha+1}{n\alpha+2} \nonumber \\
		K_{s} & = \frac{(n-1)\alpha+1}{(n-1)(n\alpha+2)}
\end{align}
\end{lemma}

From $\eqref{231}$ we see that the centrality of the hub vertex is high relative to the leaf vertices for low values of $n$.  To determine a pairwise stable condition we consider the result of two possible strategies: (1) a link is added between two leaf vertices and (2) the a leaf vertex deletes its single link with the hub vertex.

\begin{lemma}
For any single-degree vertices $i,j$ in a star graph $S_{n}$, if a link is formed between them, i.e., $a_{ij}$=1, then the centrality $K_{s}^{(+1)}$ of both vertices is given by
	\begin{equation} \label{232}
		K_{s}^{(+1)} = \frac{(n-3)\alpha^2+(1-n)\alpha-2}{(n-3)(\alpha-1)n\alpha-2n-6\alpha}
	\end{equation}
\label{lem:Lemma4}
\end{lemma}

\begin{figure}[htbp]
		\centering
\includegraphics[scale = .425]{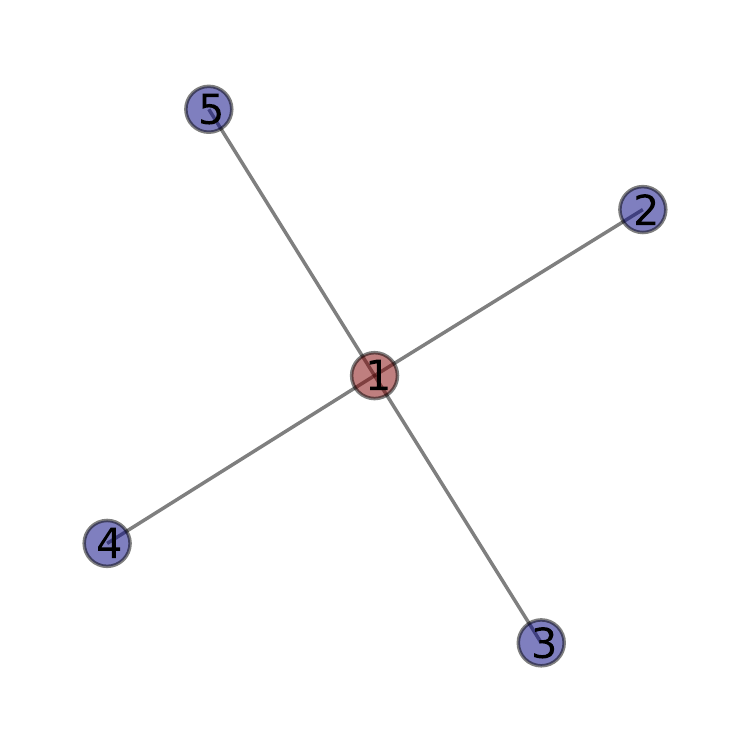}
\includegraphics[scale = .425]{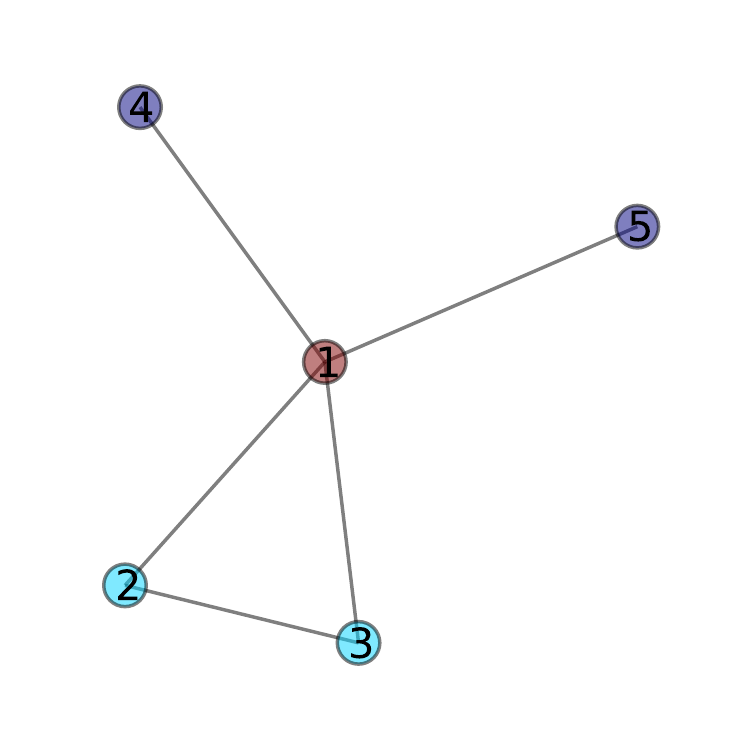}
\caption{The star graph S$_{5}$ and the graph G$_5,5$ s.t. $a_{2,3}$=1.  Star topologies arise often in many network contexts.  For lower values of $\gamma$, s$_{1}$ strategy dominates as leaf nodes connect with each other.  The resulting stable graph is often the complete graph.}
\label{fig:Star}
\end{figure}

The action in Lemma \ref{lem:Lemma4} are illustrated in Figure \ref{fig:Star}

\begin{theorem}
If the central node of a star network $S_{n}$ has a link cost $\zeta$, while the leaves have link cost $\delta$ so that $\zeta \leq \gamma$, then $S_{n}$ is pairwise stable if for every leaf $i$ with a link cost $\delta$, $R_{i}(n-1)(K_{i}^{(+1)} - K_{i}) < \delta < R_{i}(n-1)K_{i}$ and for the hub $j$ with a link cost $\zeta$, $R_{j}(n-1)K_{j}-R_{j}(n-2)K_{j}^{(-1)} > \zeta$.
\end{theorem}

\begin{remark} Before providing the proof, we remark, we will derive exact conditions: $R_{i}(n-1)(K_{i}^{(+1)} - K_{i}) < \delta < R_{i}(n-1)K_{i}$ and $R_{j}(n-1)K_{j}-R_{j}(n-2)K_{j}^{(-1)} > \zeta$.
\end{remark}

\begin{IEEEproof}
Assume that there is a payoff function for vertex $i$ defined as
\begin{equation*}
	\pi_{i} = R_{i}p_{i}K_{i}-\sum_{j\neq i}\gamma_{i}A_{ij}
\end{equation*}

\begin{equation*}
\gamma_{i} =
\begin{cases}
	\zeta & \text{if } \eta_{i}=(n-1) \\
	\delta & \text{otherwise} \\
\end{cases}
\end{equation*}
where $R_{i}$ is a constant reward obtained by vertex $i$, $p_{i}$ is the number of nodes in $i$'s component (minus itself),  $K_{i}$ is the centrality of $i$, and $\eta_{i}$ is the number degrees of vertex $i$.  Because there is only one component for a connected graph, $p_{i}=(n-1)$ for all  $i \in N$.  Let $\gamma_{i}$ be the cost for maintaining an existing link between $i$ and $j$.  In order for a star graph to be pairwise stable, the hub must have no incentive to delete its link between itself and a leaf, or
\begin{equation}
\pi_{i} - \pi_{i}^{(-1)} > 0
\end{equation}
We expand this using the centrality terms defined in $\eqref{231}$ and $\eqref{232}$ to show that a hub will not delete a link if
\begin{align}
	& \frac{(n-1)(\alpha+1)}{n\alpha+2} - \frac{(n-2)(\alpha+1)}{(n-1)\alpha+2}> \frac{\zeta}{R_{i}} \nonumber \\
	& \frac{(\alpha+1)(\alpha+2)}{(n\alpha+2)(n\alpha-\alpha+2)} > \frac{\zeta}{R_{i}}
\end{align} 
To show that a leaf will neither delete a link or establish a new link, we get these two necessary inequalities:
\begin{align}
	\pi_{i}^{(+1)}-\pi_{i} & < 0 \nonumber \\
	\pi_{i} - \pi_{iso} & > 0
\end{align}
Where $\pi_{i}^{(+1)}$ is the payoff of a leaf linking with another leaf and $\pi_{iso}$ is the payoff for an isolate node.  Because an isolated node has no neighbors, it's payoff is zero.  From this we show that any leaf $i$ will not disconnect from the hub if
\begin{equation}
	\frac{(n-1)^2\alpha+n-1}{(n-1)(n\alpha+2)} > \frac{\delta}{R_{i}}
\end{equation}
and $i$ will not link with another leaf as long as
\begin{multline}
	\frac{(n-1)((n-3)\alpha^2+(1-n)\alpha-2)}{(n-3)(\alpha-1)n\alpha-2n-6\alpha} - \\
	 \frac{(n-1)^2\alpha+n-1}{(n-1)(n\alpha+2)} < \frac{\delta}{R_{i}}
\end{multline}
Rewriting these expressions in terms of variables in the payoff function, a star graph is pairwise stable if for any leaf $i$:
\begin{equation}
	(n-1)(K_{i}^{(+1)} - K_{i}) < \frac{\delta}{R_{i}} < (n-1)K_{i}
\end{equation}
and for any hub $j$:
\begin{equation}
(n-1)K_{j}-(n-2)K_{j}^{(-1)} > \frac{\zeta}{R_{j}}
\end{equation}
\end{IEEEproof}

\section{Empirical Results}
We have shown that star networks are pairwise stable for games with heterogeneous link costs between players on a small-scale.  To illustrate the complexity of real world networks with numerous players, we have developed an algorithm that computes possible pairwise stable game solutions given a game definition and a specific set of parameters.  The blueprint of the algorithm is as follows:
\begin{enumerate*}
\item Add $N$ isolate nodes to the null network. 
\item Each isolated node is given an objective function $\pi$ and strategy set $\{s_{1},s_{2}\}$. 
\item Choose two nodes $i,j$ at random from the set $\{1,2,\ldots,N\}$.  If both (either) nodes benefit from the addition (deletion) of a link between them, the a$_{ij}$th entry in the network's adjacency matrix $\mathbf{A}$ is changed to a one (zero). 
\item A stable network is achieved when neither the addition nor deletion of a single link results in a higher payoff for all $i$ in $N$.
\end{enumerate*}
Note that the algorithm does not assume players know which links contribute a higher payoff compared to others, i.e., the interactions are treated as random events between players with imperfect information.  This is an important distinction between our model and network games proposed in previous literature \cite{HG06}.

\begin{figure}[htbp]
		\centering
		\includegraphics[scale = .25]{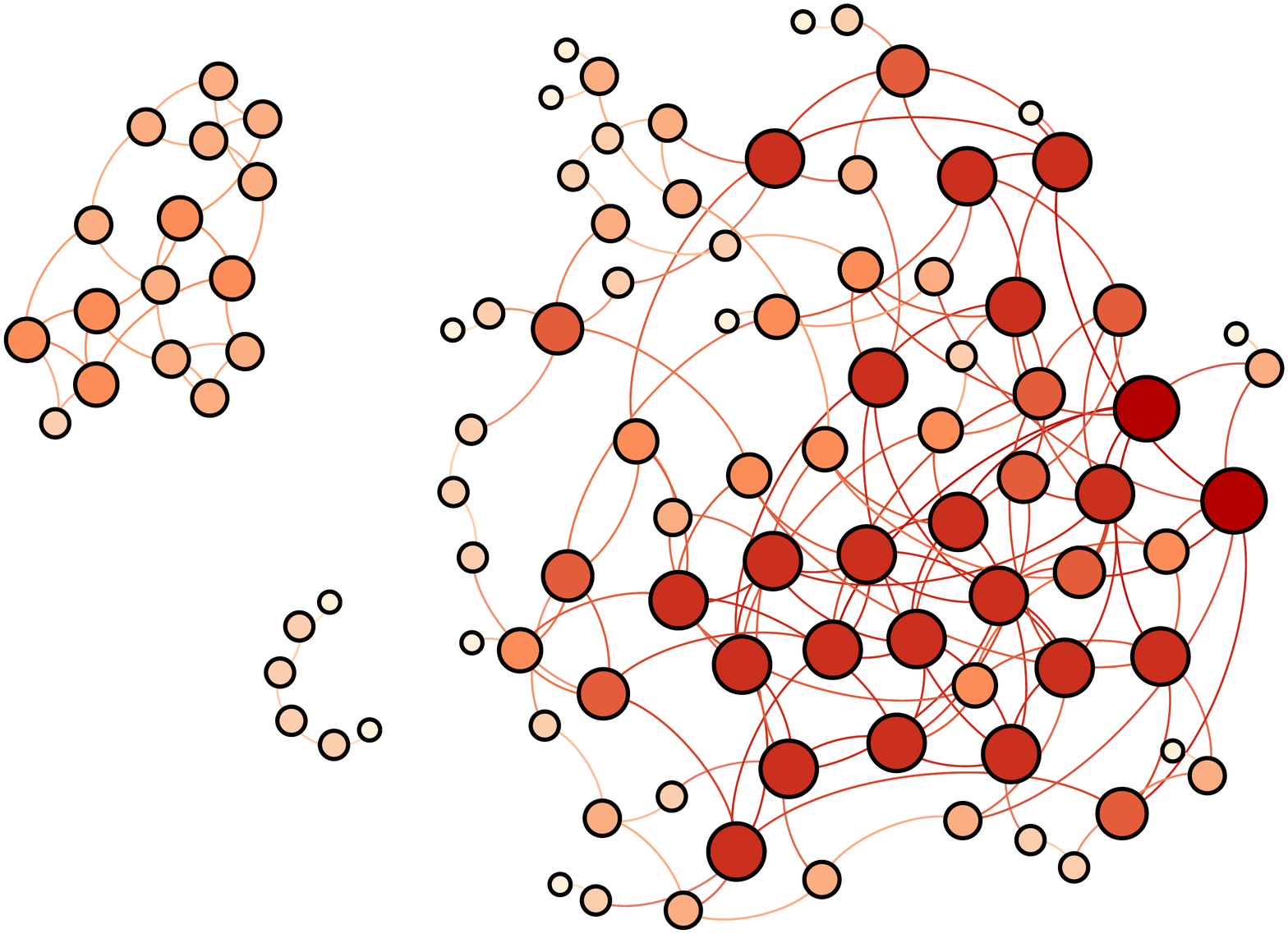}
		\caption{A simulated pairwise stable solution for the game with N=100, $\alpha$=.0075, $\gamma$=.25.  Size and color intensity is determined by the degrees of the node.}
		\label{fig:Graph1}
\end{figure}

Using this algorithm, we analyze the game $\mathcal{G}(100, S, \pi)$ with network parameters $R_{i}=1$, $\alpha=.0075$, and $\gamma=.25$,  where $\pi$ is the payoff function defined in $\eqref{pi}$.  As previously described, the player's objective relies on centralizing his position in the network at the cost of fixed-price connections.  The network shown in Fig. \ref{fig:Graph1} is one of many pairwise stable topologies resulting from these parameters.  
Notice that we have multiple connected components with a single giant component emerging from the network simulation.  This is consistent with our previous remark regarding the stability of increasingly large single component complete networks.  Without further analysis, we infer that either players in the smaller components cannot afford to be part of a larger network at the cost of their current position, or (more likely) a player in the giant component would not not benefit from taking on an additional link with a ``lesser" central player.  The pairwise stable graph shown in Fig. \ref{fig:Graph1} has an average degree of 3.56 and total payoff ($\sum_{i=1}^N \pi_{i}$) of 7.50.
\begin{figure}[htbp]
		\centering
		\includegraphics[scale = .25]{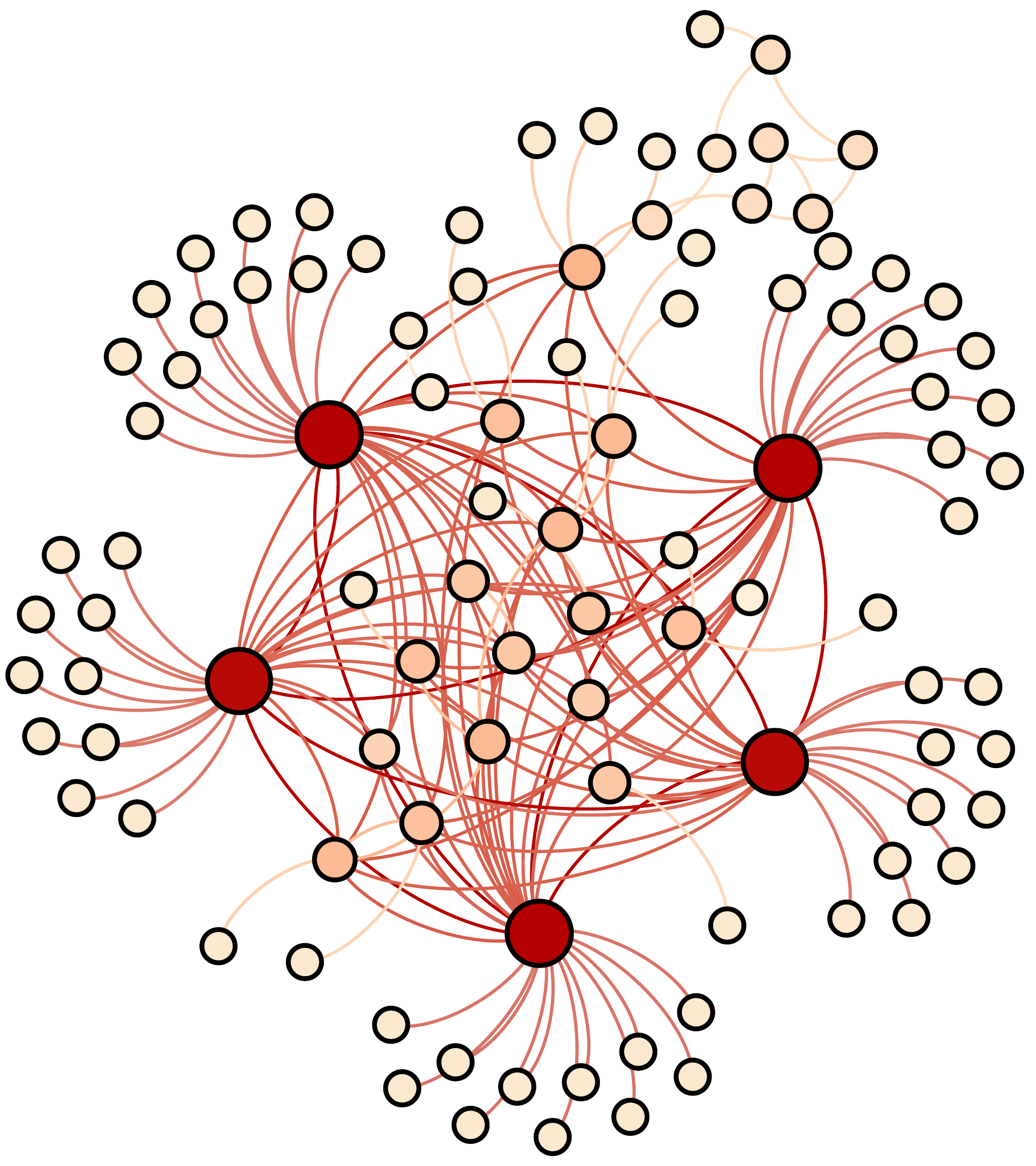}
		\caption{A simulated pairwise stable solution for the game with N=100, $\alpha$=.0075, $\delta$=.25, $\zeta$=.20.  Size and color intensity is determined by the degrees of the node.}
		\label{fig:Graph2}
\end{figure}
Next we consider an altered version of the game defined above, where a select number of players are offered a discounted link cost.  We define the new network parameters $R_{i}=1$, $\alpha=.0075$, $\delta=.25$, $\zeta=.20$, where $\zeta$ is the link cost for exactly five incentivized players.  One example pairwise stable network (shown in Fig. \ref{fig:Graph2}) exhibits an entirely different structure than the previously studied network.  The five players act as hubs, accumulating a large portion of the links to form pseudo-star topologies within the network.  Over multiple simulations, we observe that hubs only accumulate when players are offered lower link costs compared to their peers.  However, a lower link cost does not necessarily guarantee that player becomes a hub.  It appears that the number of hubs a network can sustain exhibits a positive correlation with the size of the network; intuitively, a network can only sustain a few ``superstar" nodes vying for connections.

\begin{figure}[htbp]
		\centering
		\includegraphics[scale = .375]{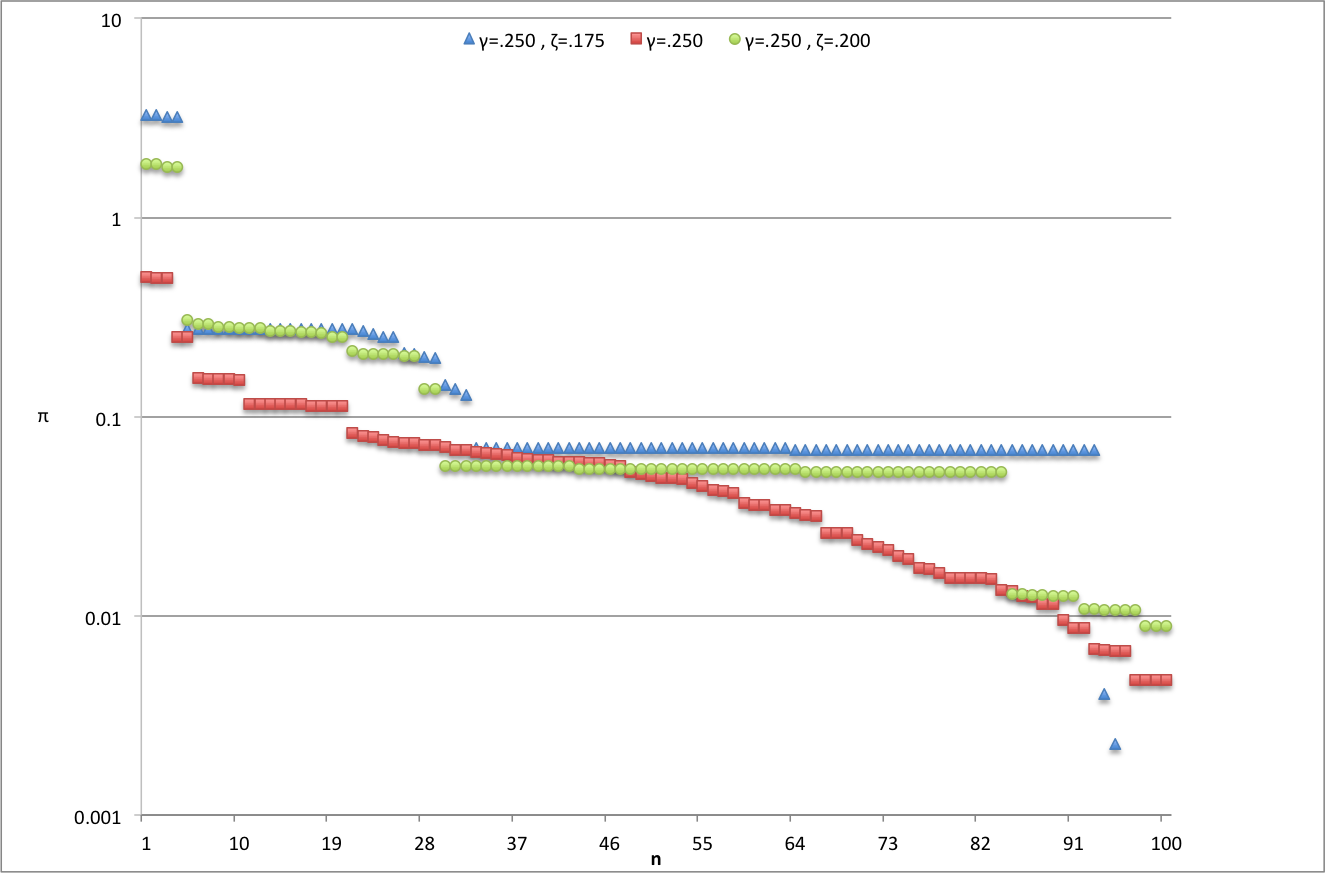}
		\caption{The payoffs of 100 players with variable link costs, shown on a base 10 logarithmic scale.  Note that as the margin between costs is increased, the payoffs of hub and intermediate players increases as well.}
		\label{fig:Runs}
\end{figure}

Interestingly, the incentivized network sustains an average degree of 3.42 and a larger total payoff of 18.45.  Incorporating incentivized players in the network not only increases the total welfare of the network, but also a number of players $without$ discounted link costs.  From Fig. \ref{fig:Runs} we see that players indirectly benefit from the centrality of neighboring hubs and are thus more highly centralized in a larger component.  Representative runs are shown in Figure \ref{fig:Runs}.

\section{Conclusions and Future Directions}

In this paper we have created an $N$-player game that models the strategic interaction between players vying for centrality within a dynamic system.  From our objective function we derived parameter-specific $(n,\alpha,\gamma)$ conditions for pairwise stability in the some commonly found real world network topologies and used those to interpret our algorithm's empirical results.  Interestingly, our research has found that introducing incentivized players with inherent ``advantages" into the model increases the total welfare of the network and creates a better semblance of communal structure when compared to a completely homogeneous population of players.  

In the future, we would like to analyze the results of more complex parameter adjustments, such as implementing an evolving cost distribution based on a player's current position within the network.  The algorithm described in the paper can further be adjusted to allow for games with perfect information, where players possess more possible linking strategies to choose from.  

In addition, we would like to analyze games that produce more realistic small-world behavior.  These networks are found in numerous social systems \cite{WT99} and a complex game theoretic model analyzing the dynamics of small-world phenomena may be of interest. 

\section*{Acknowledgment}
Portions of Dr. Griffin's work were supported by professional development
funding from the Applied Research Laboratory. Portions of Mr. Tatko's work were
supported by the Applied Research Laboratory undergraduate honors program. \newline

\bibliographystyle{IEEEtran}
\bibliography{PaperReferences,MorePaperReferences}

\end{document}